\documentclass[conference]{IEEEtran}
\usepackage{url, cite}
\usepackage{mathtools, amsmath, amsthm}
\usepackage{algorithmicx, algpseudocode}
\usepackage{graphicx,subfig, array}

\newtheorem{thm}{Theorem}
\newtheorem*{lem*}{Lemma}

\begin{document}

\title{Rules in Play: On the Complexity of Routing Tables and Firewalls}

\author{M. Wahdwa, A. Pal, A. Shah, P. Mittal and H. B. Acharya\\
IIIT Delhi, India\\
acharya@iiitd.ac.in}

\pagenumbering{arabic}
\pagestyle{plain}

\maketitle

\begin{abstract}
  A fundamental component of networking infrastructure is the
  \emph{policy}, used in routing tables and firewalls. Accordingly,
  there has been extensive study of policies. However, the theory of
  such policies indicates that the size of the decision tree for a
  policy is very large ( $O((2n)^d)$, where the policy has $n$ rules
  and examines $d$ features of packets). If this was indeed the case,
  the existing algorithms to detect anomalies, conflicts, and
  redundancies would not be tractable for practical policies (say,
  $n = 1000$ and $d = 10$). In this paper, we clear up this apparent
  paradox. Using the concept of ``rules in play,'' we calculate the
  actual upper bound on the size of the decision tree, and demonstrate
  how three other factors - narrow fields, singletons, and all-matches
  - make the problem tractable in practice. We also show how this
  concept may be used to solve an open problem: pruning a policy to
  the minimum possible number of rules, without changing its meaning.
\end{abstract}

\section{Introduction}

Policies, such as routing and filtering policies (implemented in
routing tables and firewalls) are essential to the operation of
networks. In current, packet-switched networks, routers, switches and
middleboxes examine incoming packets, and decide (based on the
relevant policy) what course of action to pursue for each packet.
Accordingly, fast packet \emph{resolution}, the operation of checking
which rule in the policy is applicable for the packet, is an important
area of research. A second important use of policies is in access
control lists, for firewalls and intrusion detection; as the security
of the system depends on the correctness of such a policy, policy
\emph{verification} is also an important problem. Algorithms for
fast operation, design, optimization, and verification of policies
work on a decision diagram representation of these policies; a 
good way to measure the complexity of a policy is the size of its
decision diagram.

However, the theory of policy decison diagrams comes with a major
caveat. The best known bound for the size of such a diagram is very
large - specifically, $(2n-1)^d$, where the policy has $n$ rules and
the number of features of interest (which we call 'fields') is $d$. As
$n$ is of the order of thousands of rules, and $d$ can easily be
$5-10$, it would seem that the complexity of these algorithms is too
large for practical cases. But this is not the case at all - in fact,
these algorithms are in common use to analyze policies and entire
networks \cite{Shaer09}!

This paper solves this apparent paradox. We deepen the theory of
policy decision diagrams with the idea of keeping track of ``rules in
play,'' which allows us to determine a tight upper bound on the size
of a decision diagram and show why the size does not explode for
practical policies. We also suggest how this allows us to optimize
policies, and obtain truly minimum (rather than just minimal, i.e.
with no redundant rules) policies.

We begin by providing our definitions and concepts, in the next
section.  Next, we discuss decision diagrams and the algorithm to
create them, as well as how we modify it to keep track of rules in
play. After this, we demonstrate how there are mitigating factors -
narrow fields, singletons, and all-matches - that reduce the size of
decision diagrams in practice. Finally, we provide our experimental
results, discuss how our paper fits into the context of related work,
and finish with some concluding remarks.

\section{Terms and Concepts}
\label{sec:Defn}

In this section, we define the terms and concepts used in the paper,
such as policies and properties, and formally introduce our concepts of
\emph{oneprob}, \emph{allprob}, and \emph{fieldwidth}. We will
introduce decision diagrams and ``rules-in-play'' in the next section.

\subsection{Packets, Rules, and Matching}

In our work, we model a \emph{packet} as a $d$-tuple of non-negative
integers. Why this model? In order to decide what to do with a packet
(whether to forward it, which interface to forward it on, etc.),
routers and firewalls examine its various attributes - usually values
in the packet header, such as source address, destination address,
source port, destination port, protocol, and so on. (In `deep packet
inspection', attributes from the packet payload are also checked.) The
$d$ fields of our packet model represent the $d$ features examined.

A \emph{rule} represents a single rule in a flow table. It consists of
two parts: a \emph{predicate} and a \emph{decision}.

The rule predicate is of the form
\[ x_1 \in [x_{1,1}, x_{1,2}] \land x_2 \in [x_{2,1}, x_{2,2}]
\ldots x_d \in [x_{d,1}, x_{d,2}] \] where each interval $[x_{k,1},
  x_{k,2}]$ is an interval of non negative integers, drawn from the
domain of field $k$. (For example, suppose the third field in packets
and rules represents source IP address. In IPv4, the domain of this
field is $[0, 2^{32} - 1]$. Then, in any rule, $ 0 \leq x_{3,1} \leq
x_{3,2} \leq 2^{32}-1 $.)

The decision is an action, such as (in a firewall) \emph{accept} or
\emph{discard}. [We call a rule with decision \emph{accept} an
``$accept$ rule'', and a rule with decision \emph{discard} a
``$discard$ rule''.]

A packet that satisfies the predicate of a rule is said to
\emph{match} the rule. For example, the packet $(1, 26, 7)$
clearly matches the rule
\[ x_1 \in [0, 108] \land x_2 \in [21, 65535] 
\land x_3 \in [7, 616] \to accept\]

\subsection{Policies and Packet Resolution}

A \emph{policy} consists of multiple rules (as defined in the previous
subsection), and a specification of which action to execute, in the
event that multiple rules match a given packet. There are two
principal methods in use to decide the precedence of matching rules.
\begin{enumerate}
\item \emph{First Match.} The rules are arranged in sequence in the
  policy, and the action of the first rule in the sequence that is
  matched by a packet is the action executed.

  This is the method usually used in firewalls.

\item \emph{Best Match.} One specific field is chosen and, out of the
  rules matched by the packet, the one with the smallest interval for
  this particular field has precedence.

  This is the method used in routing tables. The field chosen is the
  destination IP, and this technique is called 'longest prefix
  matching'.  (In routing tables, IP address intervals are usually
  denoted by prefixes, e.g., $100.150.200.0 - 100.150.200.255$, which
  is really the interval $[1687603200, 1687603455]$, is written
  $100.150.200.0/24$. Thus the best match is by the rule which had the
  longest prefix out of the rules matched by the packet.)
\end{enumerate}

The practical order of deciding precedence in a router is quite
complicated. In short,
\begin{enumerate}
\item First, find the best match.
\item In case of conflict, choose rules in the order:
  \begin{enumerate}
  \item Static routes
  \item Dynamic routes, in order \\(usually EIGRP, OSPF, ISIS, RIP)
  \item Default route
  \end{enumerate}
\item If no rule matches, discard the packet.
\end{enumerate}

However, this entire procedure can be effectively reduced to first
match, simply by ordering the rules! (The primary sort key in this
ordering is the specified prefix length, and the secondary sort key,
the type of rule.) Hence, in our work, we assume first match
semantics for policies.

The `winning' rule, the decision of which is implemented for a packet,
is said to \emph{resolve} the packet.

\subsection{OneProb, AllProb, and FieldWidth}

In policy research, the complexity of a policy is considered to be
affected by two principal factors. The first is $n$, the number of
rules in the policy, which can be up to several thousand; the other is
$d$, the number of fields in a rule, which is usually around $5 -
10$. For example, the complexity of the simple standard algorithm for
resolution - check, for each of $d$ fields of the packet, that the
value falls in the interval specified by the rule; repeat until a rule
is matched - is clearly $O(nd)$.

However, we contend that this is not sufficient information to predict
the complexity of a policy. There are two other metrics of interest. 

The first observation is that, in practice, rules are almost always
targeted to very specific uses; as a result, many of their fields
are set to either a single specific value, or to ``All'' - i.e. the
entire domain of the field. (For example, firewalls built using
Structured Firewall Design\cite{LiuGMN04} have many fields set to
All.)  As demonstrated in earlier work\cite{Windex}, policies where
high proportions of fields are set to single values, or to All, show
significantly different behavior than other policies with the same
values of $n$ and $d$. To capture this, we use the metrics $allprob$
and $oneprob$.

$allprob$ is the probability that, on randomly choosing a field and a
rule in the policy, the interval specified by the rule for the field
is ``All''.

$oneprob$ is the probability that the interval specified by a randomly
chosen rule for a randomly chosen field, is a single value, such as
$[6, 6]$.

The second observation is that the domains of different fields are of
very different sizes. For example, IPv4 addresses take $32$ bits,
protocol takes $8$ bits, and version takes $4$ bits in a standard
header. We use the metric of $fieldwidth$ to measure narrow fields
(whose domain is expressed in a small number of bits); it is the total
number of bits needed to express all narrow fields in the policy. The
reason for introducing a new metric is that narrow and wide fields
have different effects on the complexity of a policy. We provide
details in Section \ref{sec:smaller}.

\section{Decision Diagrams}
\label{sec:FDD}

In this section, we describe how policies can be represented using
Decision Diagrams, as proposed by Gouda \cite{LiuGMN04}.

A decision diagram (over fields $f_1 .. f_d$) is an acyclic, rooted
digraph. Every \emph{terminal} node, i.e. one with no outgoing edges,
is marked with a decision (for example, in case of firewalls,
decisions are \emph{accept} or \emph{discard}). Every non-terminal
node is marked with the name of a field, and has outgoing edges marked
with an interval of values for its field; values marked on the edges
from a node do not overlap.

There is at least one directed path from the root to every other node,
and no directed path has more than one node labeled with the same
field.  Thus, a decision diagram represents a simple deterministic
finite automaton. Given any packet $ p = (p.f_1, .. p.f_d )$, there is
exactly one path from the root corresponding to the field values of
$p$, and it terminates in a terminal node; the decision at this node
is the decision of the policy for $p$.

We consider the decision diagram in its fully expanded form, as a
tree. In our example, the root is labeled $f_1$, its children are
$f_2$, down to $f_d$, whose children are the leaf i.e. terminal nodes.

To resolve a packet, we follow a path from the root down to a leaf
node, at each node choosing the edge matched by the packet - i.e.,
labeled with the interval containing the value of the corresponding 
field for the packet. Given a policy with $n$ rules of $d$ fields, a 
decision diagram performs packet resolution in $O(d)$ time. Policy 
verification has the same time complexity as the size of the diagram. 
(It is essential to check the decision corresponding to packets that 
match the property; in the worst case , the property specifies all 
possible packets, and it is necessary to check paths from the root 
to every terminal node.)

\begin{figure*}
\caption{Example: building a decision diagram, step by step.}
\label{fig:1}
\begin{tabular}{cc}
\subfloat{\includegraphics[height=1.5in]{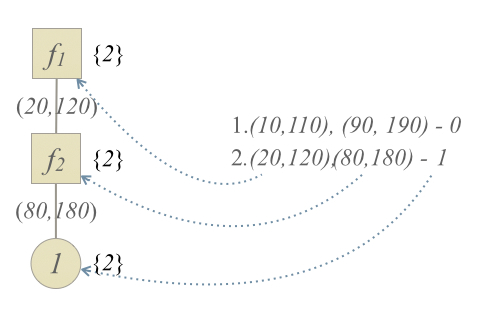}} &
\subfloat{\includegraphics[height=3.5in]{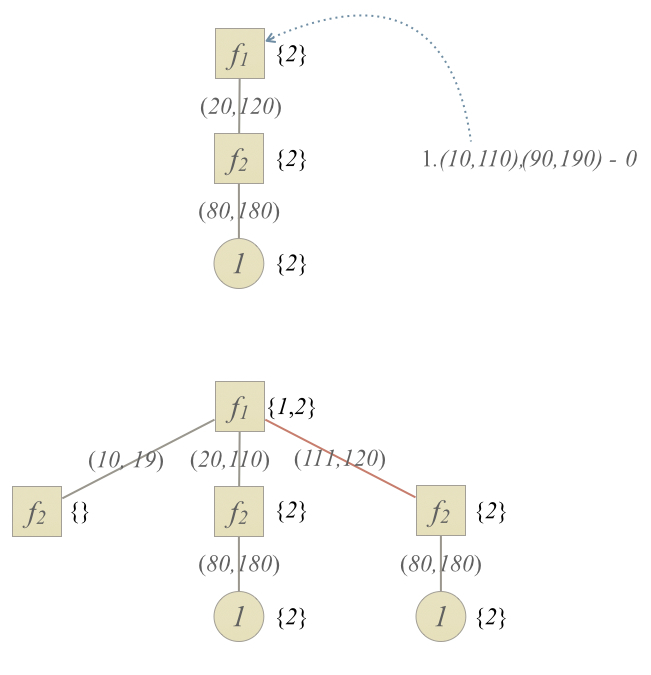}}\\
\subfloat{\text{Step $1$. First rule from bottom, i.e. rule $2$}} &
\subfloat{\text{Step $2$. Adding $f_1$ from next rule, rule $1$.} }\\
\subfloat{\includegraphics[height=3.5in]{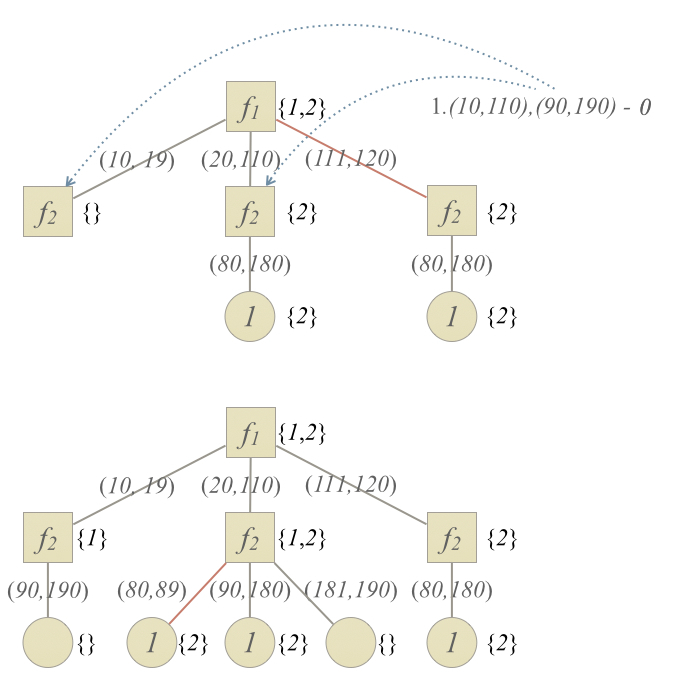}} &
\subfloat{\includegraphics[height=3.5in]{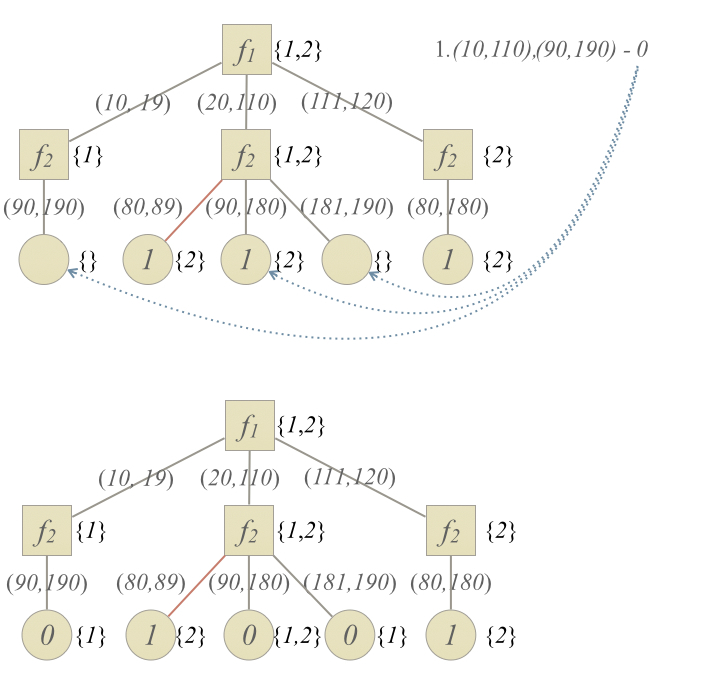}} \\
\subfloat{\text{Step $3$. Adding $f_2$ from rule $1$.}} &
\subfloat{\text{Step $4$. Adding decision from rule $1$.}} 
\end{tabular}
\end{figure*}

We now extend the algorithm to construct decision diagrams, by
introducing the concept of \emph{rules in play}. A rule $R$ of a
policy is ``in play'' at a node if at least one packet matching $R$
follows a path through the node when it is resolved. $R$ is in play at
an edge if it is in play at the node into which the edge leads.
Annotating the nodes with the rules in play requires only slight
modification to the algorithm for building a decision diagram; we
present the algorithm in Figure \ref{alg:1}. In Figure \ref{fig:1}, we
show an example, building an annotated decision diagram from the
policy
\begin{align*}
x \in [10, 110] \land y \in [90, 190] \to 0 \\
x \in [20, 120] \land y \in [80, 180] \to 1 \\
\end{align*}

For our first demonstration of how rules in play can be useful, we
demonstrate how, given a policy, we can use the rules in play at the
leaf nodes, and find a $minimum$ equivalent policy (equivalent
policies have the same decision for every packet).  We delete rules,
without changing the decision of the policy for the packets reaching
any leaf node. (As we are not adding rules, there will be no new paths
or new leaves to consider.)

Consider a leaf node with the $accept$ rules $3, 13$ and the $discard$
rule $7$ in play. So long as rule $3$ remains, deleting other rules
does not affect the decision at this leaf. What if we delete rule $3$?
We must delete rule $7$ (so it does not become the first rule to match
these packets), and keep rule $13$.  Using $x_i$ to mean that rule $i$
is present, we must satisfy
\[ x_3 \lor (\lnot x_7 \land (x_{13}) ) \]
We can apply this logic recursively:
if we had $accept$ rules $3$, $13$ and $23$, and $discard$ rules $7$,
$10$ and $17$, then
\[ x_3 \lor ( \lnot x_7 \land (\lnot x_{10} \land ( x_{13} \lor ( \lnot x_{17} \land
(x_{23}))))) \]
must be satisfied, to preserve the policy decision at this leaf.

\begin{enumerate}
\item At a leaf node, we start with the first rule in play. All other
  rules with the same decision are \emph{complying} rules, and 
  the others are \emph{conflicting} rules.
\item We traverse the list of rules in play, in order, building an 
expression (string). 
\item If the rule $R_i$ is complying, we add ``$x_i \lor ( $''. If it is
conflicting, we add ``$\lnot x_i \land ($''.
\item At the end, we close all parentheses.
\end{enumerate}
Taking the AND of the expressions for all the leaf nodes, we get an
expression which must be satisfied to keep the decision of the policy
unchanged for all packets.

To find the minimum policy, we create our expression for the original
policy (as described above), and search for the solution with the
smallest number of $x_i$ set to $1$ - a Min-One SAT problem. This
gives us the smallest combination of the rules in the policy that
preserve its semantics for all packets, a minimum policy (rather than
the minimal policies obtained by trimming as many redundant rules as
possible). So far, this algorithm is too slow to be practical for
policies of more than $50$ rules; however, we expect optimizing the
solver will make it fast enough for real use. We will develop this
further in future work.

In the next two sections, we show how, by keeping track of rules in
play, we can prove that decision diagrams do not grow exponentially 
large in practice.
\begin{figure}
\begin{small}
\caption{Building annotated decision diagram from policy}
\label{alg:1}
\begin{algorithmic}
\Procedure{AddnewRule}{Rule $R$, name $R.name$, node $x$} 
\State \textbf{Add} $R.name$ to list of rules in play at $x$
\If {$x$ is a terminal node}
\State \textbf{Label} $x$ with the decision of $R$.
\Else
\State $i$ is the field of node $x$.
\State $R.i$ is the interval for field $i$ in $R$.
\State $x.i_1, x.i_2 ..$ are the values on
the edges from node $x$.
\State \textbf{Build} new paths from $x$, starting with new 
outgoing edges labeled with the intervals in $R.i - \{x.i_1, x.i_2
.. \}$.
\Comment To each edge, add new nodes and edge labels as per
remaining fields in $R$.  The terminal node has decision of $R$. 
These nodes start with only one rule in play, i.e. $R.name$.
\For{all labels $x.i_k$}
\If {$x.i_k \cap R.i$ is empty} 
\State \textbf{continue}
\Else
\State $y$ is the target of edge labeled $x.i_k$.
\For{every interval $x.i_{new} \in x.i_k - R.i$}
\State \textbf{Copy} (subgraph rooted at $y$ ).
\State $y'$ is the new copy of $y$ itself.
\State \textbf{Add} edge $x \to y'$, labeled $x.i_{new}$.
\EndFor
\State \textbf{Relabel} edge $x \to y$ with $x.i_k \cap R.i$. 
\State $AddNewRule(R, R.name, y)$
\EndIf
\EndFor
\EndIf
\EndProcedure
\end{algorithmic}

\begin{algorithmic}
\Procedure{BuildDDiagram}{Policy $\{R_1, R_2 .. R_n\}$} 
\State \textbf{Create} empty node $x$ with no outgoing edges.
\State \textbf{Label} $x$ with desired field for root.
\For{$index \gets n ... 1,$ step $-1$}
\State $AddNewRule(R_{index}, index, x)$
\EndFor
\State \textbf{return} Decision Diagram rooted at $x$.
\EndProcedure
\end{algorithmic}
\end{small}
\end{figure}

\section{The Size of Decision Diagrams}

In this section, we determine the worst case size of a policy decision
diagram, measured as the number of leaf nodes of the diagram. We begin
with the existing upper bound.

The intuition behind the bound is that, as the decision diagram is a
tree of fixed depth, its size can be maximized by making the branching
factor of each node as large as possible.

As there are $n$ rules in the policy, there can be a maximum of
$2n - 1$ outgoing edges for a node. (Each edge must be labeled with at
least one interval. The $n$ rules have $2n$ end points - each interval
has a start and an end. These $2n$ end points thus divide the domain
into a maximum of $2n - 1$ intervals. Hence we can have at most
$2n - 1$ edges from a node.)

Thus, an upper bound on the size of the decision diagram for a policy
of $n$ rules and $d$ fields is $(2n - 1)^d$.

Our construction, which is mindful of rules in play, allows us to
tighten this bound considerably. The intuition behind this is that,
while a node with $n$ rules in play can indeed have a branching factor
of $(2n - 1)$, as seen above, not all these branches still have $n$
rules in play! In fact, of the outgoing edges, exactly \emph{one} edge
still has all $n$ rules in play (the one corresponding to the
intersection of all the rules). There are also edges with $n-1$ rules
in play, $n-2$ rules in play, and so on, down to edges with $1$ rule
in play. In our example from Figure \ref{fig:1}, consider the edges
emerging from the root: edge $(10, 19)$ has only one rule in play (the
first), edge $(20, 110)$ has both rules in play, and edge $(111, 120)$
has only one rule in play (the second).

So, in order to construct the largest decision diagram, there are two
factors to maximize:
\begin{enumerate}
\item The number of branches, i.e. the branching factor at each node
\item The number of rules in play along each branch.
\end{enumerate}

In fact, in the largest possible decision diagram for a given $n$ and
$d$, the pattern is that one outgoing edge has $n$ rules in play, two
have $n-1$, two have $n-2$, and so on, down to the last two edges
which have a single rule in play. This is a dynamic programming
problem: given the largest decision diagrams that can be built with
$1, 2 ... n$ rules in play, at a depth $d-1$, we can compute the size
of the largest decision diagram with depth $d$, and $n$ rules in play.
We also have the following observations.
\begin{enumerate}
\item For any $d$, when $n=1$, the number of leaves is $1$. 
(Only a single rule is in play, so there is no branching.)
\item $d=0$ indicates we are at a leaf node, so the size is $1$.
\item When $d = 1$, and the number of rules in play is $n$, then the 
size of the decision diagram (measured by number of leaves) is $2n-1$. 
\end{enumerate}

Stating this result formally,
\begin{thm}
If we represent the maximum size (measured by number of leaf nodes) of a 
decision diagram with $n$ rules and $d$ fields as $f(n, d)$,
\begin{align*}
f(n, d) &= 2 \sum\limits_{i=1}^{n-1} f(i, d-1) + f(n, d-1) \\
& \text{for $n > 1, d > 1$} \\
f(n, 1) &= 2n-1\\
f(1, d) &= 1
\end{align*}
\end{thm}

Next, we need to prove that this intuitively appealing construction ( all
$n$ rules overlapping for one central edge, and rules ``falling off'' one by 
one to either side) is in fact the largest possible decision diagram for a
given $n$ and $d$. Our proof is as follows.

Consider any decision diagram. It forms a tree, in which every non-leaf node 
represents a field. Every edge is labeled with an interval (of values for the field
represented by the node it starts from), and associated with a list of rules in 
play at that edge.

We consider how, in constructing the old bound, we maximize the
branching factor at a node. Every rule $R_i$ has a range
$[start_i, end_i]$; these end points ($start_i$ and $end_i$) segment
the domain space into intervals. $n$ rules introduce $2n$ end points,
and each point added has the potential to create a new edge. (The set
of rules in play is different before and after an end point.)  To
maximize the number of edges, we assume that no two end points
coincide; this gives us $2n-1$ as the number of outgoing edges from a
node.

Now, we extend this argument. We will abuse the variable names $n$ and
$d$. In general, we now use $n$ to mean the number of rules in play at
a node (or an edge) ; the original meaning of $n$, the number of rules
in the policy, is the $n$ seen at the root. Similarly, we use $d$ to
mean the depth of the subtree rooted at a node; the original meaning
of $d$, the number of fields, is the $d$ seen at the root. Clearly, $f(n,d)$
increases monotonically with $n$ and with $d$.

Let us arrange the outgoing edges from a node in ascending order of
the values of their labels (intervals). [As these intervals are non
overlapping, we can sort them by their start, i.e. smallest values, or
by end, i.e. largest values - the order will be the same.] Two edges
are said to be \emph{adjacent} if there exists no edge whose label
(interval) lies in between their labels.

\begin{lem*}
  In the largest decision diagram for a given $n$ and $d$, the sets of
  rules in play on adjacent edges differ by one rule exactly.
\end{lem*}
\emph{Proof.}
Let the rulesets on two adjacent edges be $A$ and $B$.

Let $A = X + Y$ and $B = Y + Z$.

If $X$ and $Z$ are both non-null, then by increasing the $end_i$ of
rules in $X$, by $1$ (or reducing the $start_i$ of rules in $Y$, by
$1$) we can add a new edge with the ruleset $X + Y + Z$ in between
these edges.

This edge could not already exist. (Let the field be named $f_1$.
Without loss of generality, assume $f_1$ increases going from $A$ to
$B$.  The labels of rules are intervals - they start once and stop
once exactly.  $X$ appears on no edge with greater $f_1$ than edge
labeled $A$, and $Z$ on no edge with less $f_1$ than edge labeled
$B$. So the edge with $X+Y+Z$ could not have been at some other
position.)

Also, adding this edge does not affect the rest of the diagram (for
$f_1$, the other edges have different values, and are not affected;
the other fields are not touched at all). Adding this new edge clearly
increases the decision diagram size. But this is the \emph{largest}
decision diagram for this $n$ and $d$, so we have a contradiction.

Now, we have proved that either $X$ or $Z$ is the null set. Without
loss of generality, let us assume that $X$ was null.

Can $Z$ also be a null set? This is not possible, because then $A==B$!
The edges labelled $A$ and $B$ would be merged to form a single edge.

On the other hand, can $Z$ contain more than $1$ rule? 

In this case, $ Z = Z_1 + Z_2$, where $Z_1$ has one rule and $Z_2$ is 
non-empty. The $start_i$ of rules in $Z_1$ and $Z_2$ coincide. 

If we had a decision diagram where all else is the same, but these
rules did not share the same $start_i$, we would have an extra edge
labeled with $Y + Z_1$ between these two edges. But this is the
largest possible decision diagram for given $n$ and $d$, so again we
have a contradiction.

Hence, one of $A - B$ or $B - A$ is the null set, and the other has
exactly one rule. This suffices to prove our lemma.  
$\qedsymbol$

Next, we consider the first edge (i.e. the one with the smallest
$start_i$. This edge has only one rule in play; if it had more, by a
similar construction as above, we could add a preceding edge. Now the
next edge can have at most two rules in play, and if we continue to
count up, we eventually reach $n$ rules in play. After we reach $n$
rules, on the next edge we can only have $n-1$ rules in play, and so
on.

To complete our proof that the largest decision diagram indeed follows
this structure, we introduce the idea of ``closed'' rules. At every
edge, besides the number of rules in play, we can keep count of $n'$,
the number of rules for which we have already exceeded the
$end_i$. These rules can therefore not come into play at a later edge
- they are used up. In our running example, consider the last edge
$[111, 120]$ coming out of the root; this edge has the second rule in
play, and the first rule closed.

Suppose we characterize decision diagrams, for a given $n$ and $d$, by
drawing a plot, showing the number of rules in play on an edge vs. the
position of the edge (first edge is the one with smallest $start_i$,
etc.) We know that, for the largest decision diagram, adjacent
$y$-values differ by exactly $1$.  Given the monotonic nature of $f$,
clearly, if the figure for decision diagram $D1$ is enclosed within
the figure for another decision diagram $D2$, then $D2$ is larger than
$D1$.

Our construction produces a symmetric triangle sloping from $1$ up to
$n$ and from $n$ back down to $1$.  It is easy to see that, for the
rising half, the plot for the largest decision diagram is contained
within this triangle; it increases the number of rules in play as
quickly as possible.  However, it is not so obvious that this
condition also holds for the second (falling) half (because our
construction also grows the number of closed rules as slowly as
possible). Could there be a larger decision diagram whose plot
``breaks out'' from the triangle? The two plots must intersect, so
suppose it intersects the falling half after $k_o$ rules have been
brought into play and $k_c$ rules have been closed,
\begin{align*}
k_o + k_c = k \\
k_o - k_c = 2n - k \\
k_c \leq k_o \leq n 
\end{align*}
which reduces to $k_o = n, k_c = 0$, i.e. the only plot that satisfies
the given constraints is our original construction.

This concludes our proof that the largest decision diagram for a given
$n$ and $d$ indeed follows the pattern where edges have
$1, 2 ... n-1, n, n-1, ... 2, 1$ rules in play. Our bound is tight:
the following policy has exactly this decision diagram, and hence,
matches the size of this upper bound.
\begin{align*}
x_1 \in [1, 2n-1] \land ... \land x_d \in [1, 2n-1] \to accept \\
x_1 \in [2, 2n-2] \land ... \land x_d \in [2, 2n-2] \to discard \\
... \\
x_1 \in [n, n] \land ... x_d \in [n, n] \to discard
\end{align*}

\section{Size in real Decision Diagrams}
\label{sec:smaller}

In the previous section, we present our new bound for the size 
of decision diagrams. This bound is much smaller than the old
bound of $(2n-1)^d$; for example, for $n = 1000$ and $d = 10$,
\[f(n, d) = 2.808 \times 10^{26} << (2n -1)^d = 1.019 \times 10^{33}\].
However, this bound, though tight, is still quite large. In this section,
we answer the question of why, in practice, decision diagrams do not 
grow intractably large. The answer makes use of our concepts of oneprob,
allprob, and narrow fields.

\subsection{OneProb and Singletons}

In constructing a decision diagram, we have two main operations that 
increase its size. One is adding a new path; this happens when the new 
rule specifies new values for a field, i.e. values for which no outgoing 
edges exist. The other operation is when the interval specified by the 
rule only partly overlaps with the label of an edge; in this case, we 
`split' the edge and make copies of the subgraph below. (In our example, 
the rule has $x \in [10, 110] $ and the old edge has $x \in [20, 120]$.
As they partially overlap, we get new edges labeled $x \in [10, 19]$, 
$x \in [20, 110]$, and $x \in [111, 120]$.)

Now we consider the impact of \emph{singletons}. A rule is called a
singleton for field $x$ if matches only packets with one single value
of $x$, i.e. its interval for $x$ is a single value like $[29, 29]$.

Clearly, singletons cannot have partial overlap with any other rule.
As a consequence, an edge (from a node labeled $x$) with a singleton 
(for $x$) cannot be split! This leads immediately to the observation that
if all the rules are singletons, the maximum branching factor at a node
is not $2n-1$, but $n$ (when all the singletons have different values for
$x$, and thus produce different branches). However, this attractive idea
is limited in its power: as soon as we introduce a single non-singleton 
rule $R'$, we are back to $2n - 1$. [The worst case is when all $n-1$ 
singletons split $R'$, cutting up its interval into $n-1+1 = n$ parts in
between them. Now we have $n-1$ edges with two rules in play ($R'$ 
and one of the singletons), and $n$ edges with only $R'$ in play, so 
$2n-1$ edges in all. For example, consider rules with intervals $[1, 10]$,
$[3,3]$ and $[7, 7]$; we have $2 \times 3 - 1 = 5$ edges, labeled with
$[1, 2]$, $[3, 3]$, $[4, 6]$, $[7, 7]$, and $[8, 10]$. ]

To make a stronger argument, we consider the number of rules in play
along an edge.

When we have mutliple singletons for a field, there are two choices. 
They either overlap completely, or they do not overlap at all. If they 
do not overlap, this increases the branching factor at the node (more
outgoing edges). But this means the number of rules in play along 
those edges (and therefore, at the nodes the next layer down) is 
reduced! For the largest decision diagram, we need to trade off the
``immediate'' branching factor at the node and the ``potential'' for
more branching at lower levels. The solution is only obvious at nodes
where $d = 1$, as the only lower level is the leaf nodes (i.e. no more
branching is possible at lower levels); in the largest decision diagram, 
all singletons of the field where $d=1$ have distinct values. 

Singletons increase the size of the decision diagram most effectively 
if they intersect the interval with the largest possible number of rules 
in play. Consider the diagram made with the non-singleton rules; the 
edges follow the standard pattern, with $1, 2 ... n, n-1, ... 1$ rules in 
play. Now we introduce a singleton. If it splits an edge with $r$ rules 
in play, we replace one edge with $x$ rules with two edges with $r$ 
rules (before and after the singleton), plus one edge with $r+1$ rules 
(where the $r$ rules and the singleton overlap). If on the other hand, 
the singleton did not split any edge, it would introduce only one new 
edge with a single rule in play (the singleton). Clearly, the maximum 
is when $r = n$. 

How are we to proceed when introducing more singletons? This is not
straightforward. Let us assume that the number of singletons is $s$,
the number of non singletons is $t$, and the depth of the tree (below
the node we are considering) is, as usual, $d$. Then to find the size
of the largest decision diagram, we try all the partitions of $s$ singletons;
for example, if $s = 4$, we need to try $1+1+1+1, 1+2+1, 1+3, 2+2$, 
and $4$. Then, using $g(s+t, d)$ as the formula for maximum size, and 
$a$ as the number of pieces in the partition $\{s_1, s_2, ... s_a\}$,
\begin{align*}
  g(n, d) = & 2 \sum\limits_{i=1}^{t-1} f(i, d-1) \\
            & + max \Big[ (a + 1) g(t, d-1) \\
            & +  \sum\limits_{j=1}^a g(t + s_j , d-1) \Big]\\
n = & s + t
\end{align*}
where the max is taken over all partitions.

There is a small gap in this argument, which we now explain. How do 
we know the largest decision diagram is produced by adding singletons 
to the largest diagram without singletons? The reason is the monotonicity 
of the size function. With or without singletons, a diagram with more edges
and more rules in play along an edge is larger. [It is also of interest to note
that a rule being a singleton for one field does not mean it is a singleton for
the fields at lower levels! Counted as a rule in play, it is just as powerful as
any other rule.] The largest diagram to begin with, also produces the largest 
increases: other structures also grow fastest by adding singletons that split 
the edge with the most rules in play, and they cannot have edges with more
than $t$ rules in play. 
For example, consider two diagrams with the number 
of rules-in-play, $(1, 2, 1, 2, 1)$ and $(1, 2, 3, 2, 1)$. If we now add two
singletons, the first will yield a diagram with the maximum size
\[3 g(1, d) + g(2, d) + max \big( g(4,d) + 2g(2, d), 2g(3, d) + 3g(2, d) \big) \]
and the second,
\[2 g(1, d) + 2g(2, d) + max \big( g(5,d) + 2g(3, d), 2g(4, d) + 3g(3, d) \big)\]
which is clearly larger.

The probability that a rule is a singleton is given by $oneprob$, so
our intuition is that a high value of $oneprob$ leads to a smaller
decision diagram. One point that we have not highlighted, is that
singletons are also much less likely to overlap than other rules are;
as our focus is on the worst case, we could not make use of this
factor, but it most likely also plays a role in the tractable size of
practical decision diagrams.

\subsection{AllProb and All-Matches}

The behavior of \emph{all-matches}, that is, rules that match all values 
for a given field, is much less complicated. At a given node, there is no
choice about where to place the all-matches; they cover the entire domain
for the field. So the all-matches all behave like one single large rule, which
adds $u$ (the number of all-match rules) to the number of rules in play
on each edge. We can update the formula for largest decision diagram 
as follows.

\begin{align*}
  g(n, d) = & 2 \sum\limits_{i=0}^{t-1} f(i + u, d-1) \\
            & + max \Big[ (a + 1) g(t + u, d-1) \\
            & +  \sum\limits_{j=1}^a g(t + s_j + u, d-1) \Big]\\
n = & s + t + u
\end{align*}

where, again, $a$ is the number of pieces the singletons are partitioned
into, and the max is taken over all such partitions.

The probability that a rule is an all-match is given by $allprob$. An
intuitive way to explain of the effect of all-matches is that, when
$allprob$ grows sufficiently large,  as most of the rules have total
(rather than partial) overlap, the size of the decision diagram is reduced. 
In fact, in the extreme case of $allprob=100$, the decision diagram is
a linked list (from the root to the decision of the first rule, which resolves 
all packets).

\subsection{Narrow Fields}

Our construction of decision diagrams follows an implicit assumption.
When we state that the outgoing edges from a node can carve the domain
into $2n-1$ pieces, etc., we assume that the domain is large enough to 
divide into so many distinct pieces. 

However, this assumption is not true for some domains. These are the
ones we call \emph{narrow} fields. 

Clearly, if a field is represented with $w$ bits, the greatest
possible number of outgoing edges is upper bounded by $2^w$. [Every
possible value of the field labels a distinct edge.] We can extend
this argument by placing all the narrow fields together, as the fields
closest to the leaves (i.e. with $d = 1, 2 ... d_{narrow}$.  Suppose
$fieldwidth = W$ bits, i.e. the narrow fields can all be represented
using $W$ bits. Our computation for the size of the decision diagram
does not change for the non-narrow fields; however, the ``leaf'' nodes
for this tree are now the roots of decision trees for the narrow
fields, which expand to $2^W$ bits each. So we can simply compute the
size for the non-narrow bits, and multiply the result by $2^W$ to get
an upper bound on the size.  This idea is similar to the automata size
bound of Erradi \cite{erradi14}.
\begin{align*}
  g(n, d) = & 2 \sum\limits_{i=0}^{t-1} f(i + u, d-1) \\
            & + max \Big[ (a + 1) g(t + u, d-1)  \\
            & + \sum\limits_{j=1}^a g(t + s_j + u, d-1) \Big]\\
n = & s + t + u
\end{align*}
for $d> d_{narrow}$, the number of narrow fields, and
\begin{align*}
  g(n, d) = & g(n, d-1) * 2^{w_d}
\end{align*}
otherwise. ($w_d$ = bit width of field at height $d$.)
 
However, this is no longer a \emph{strict} upper bound. The reason is that,
if we do organize the decision diagram with wide fields closer to the root 
and narrow fields closer to the leaves, some of the edges from the upper
nodes to the lower nodes have a very small number of rules in play - 
$1$, $2$ etc. Clearly, for these edges, the lower nodes are no longer 
``narrow'' fields, as the bound of $2n - 1$ is again smaller than $2^w$.

Though this bound is no longer tight, in the presence of narrow fields,
it also serves to reduce the size of the decision diagram considerably.
We present our experimental results in the next section.

\section{Results}

\begin{figure*}
\caption{Performance of size bounds for Decision Diagrams}
\label{fig:2}
\subfloat[First bound: smaller than $(2n-1)^d$]{\includegraphics[width = 3.5in]{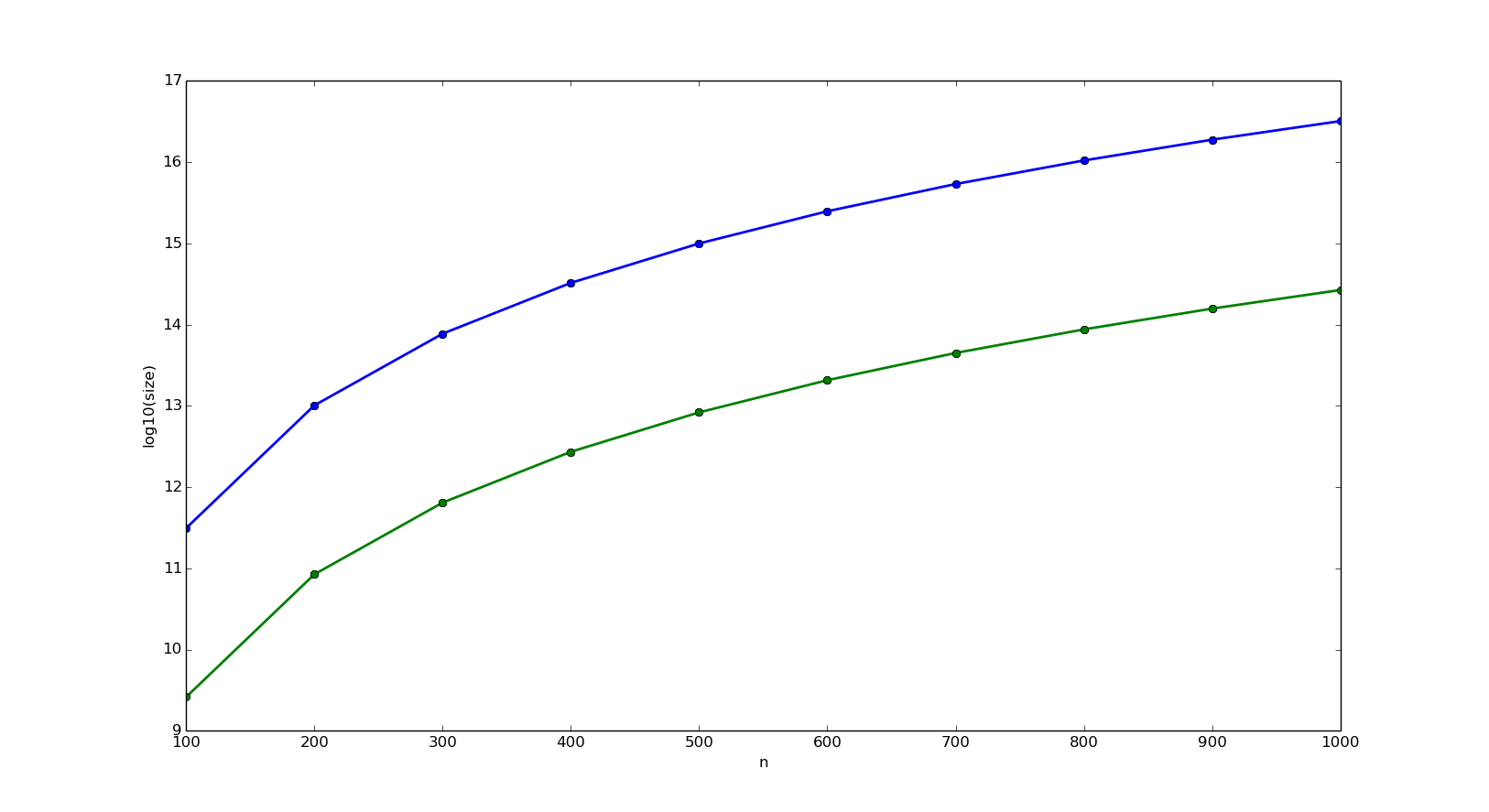}}
\qquad
\subfloat[Effect of $allprob$. (Top to bottom, $0, 20, 40, 60, 80$.)]{\includegraphics[width = 3.5in]{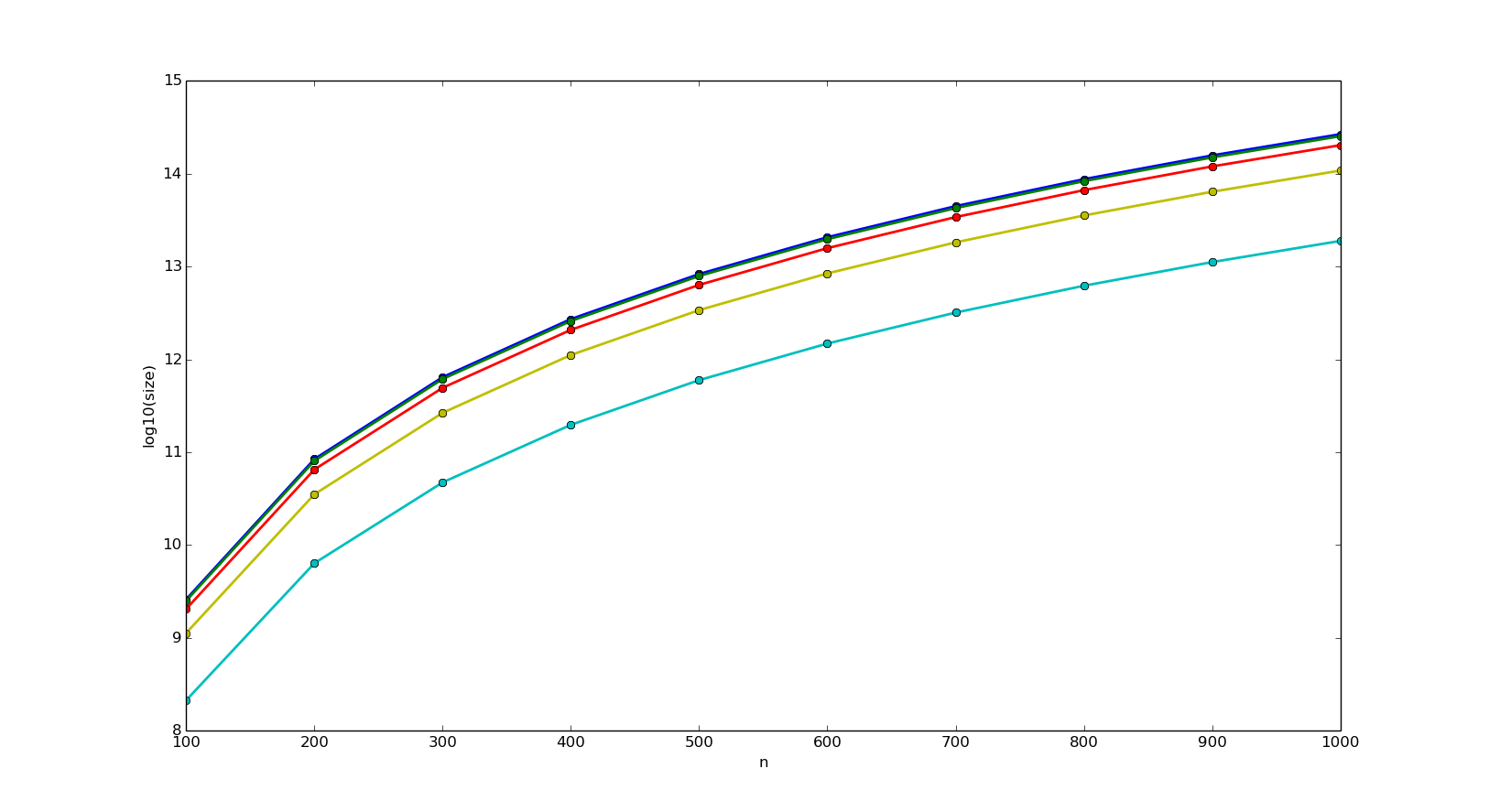}}\\
\subfloat[Effect of narrow fields. $allprob = 40$. (Lines, top to bottom, 
$5$ wide fields; $4$ wide + $1$ narrow; and $3$ wide + $2$ narrow. 
Narrow fields $2$ bits wide.)]{\includegraphics[width = 3.5in]{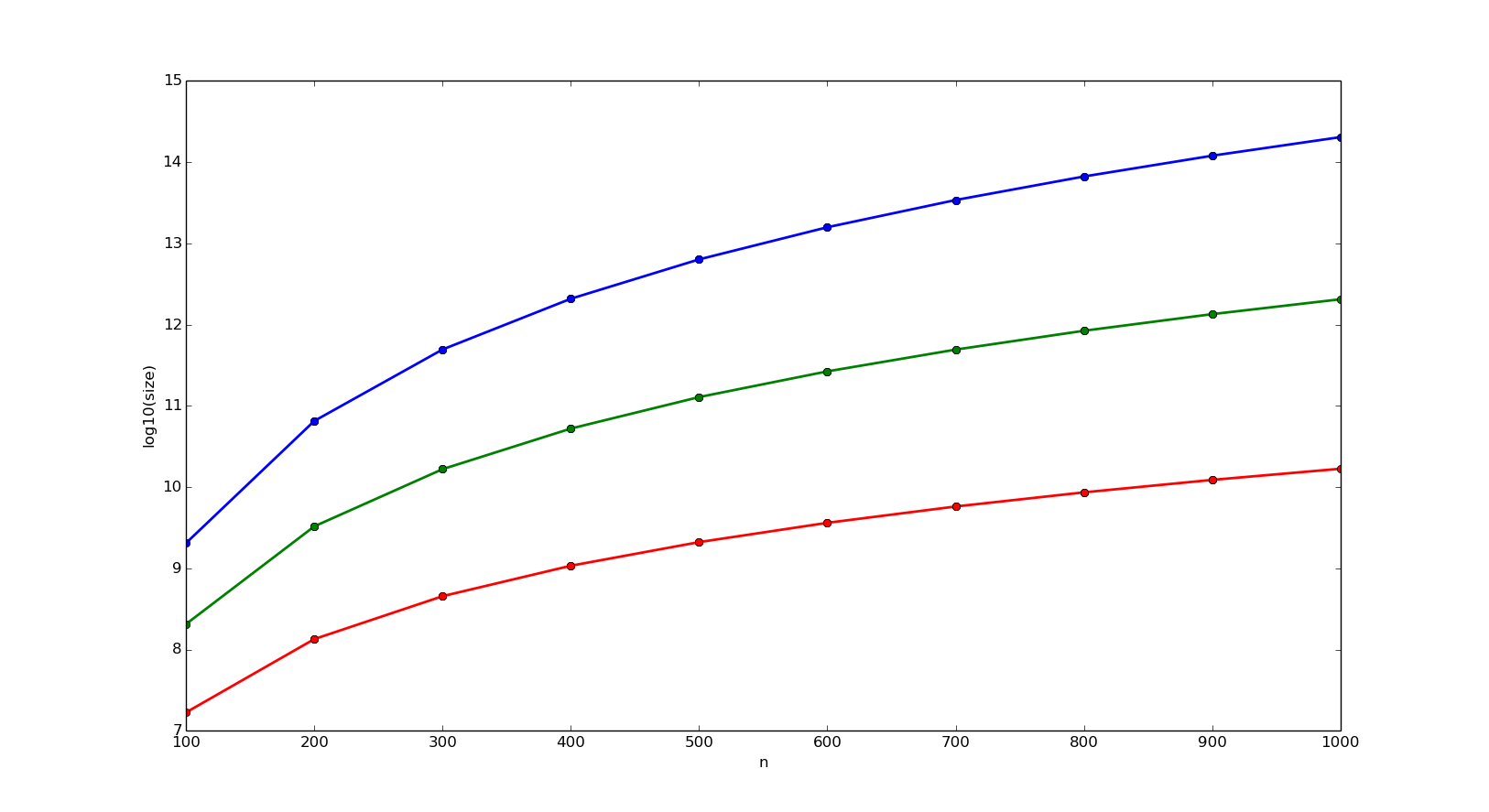}}
\qquad
\subfloat[Effect of $oneprob$. (Top to bottom, $0, 20, 40, 60$.)
$allprob = 40$, $4$ wide + $1$ narrow field. ]{\includegraphics[width = 3.5in]{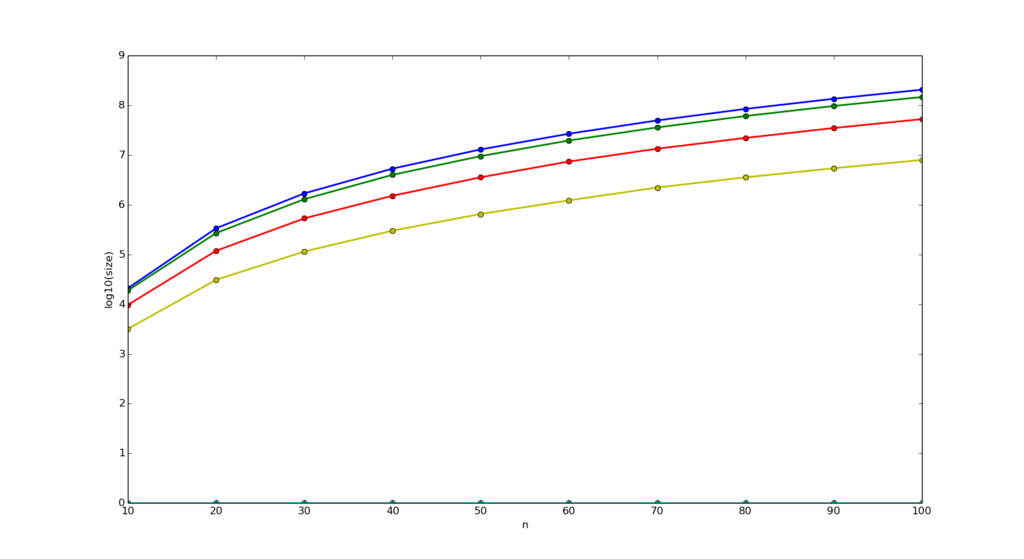}}
\end{figure*}

This section describes our experimental verification of the size
bounds for decision diagrams. In Figure \ref{fig:2}, we demonstrate
the effect of the mitigating factors, on practical policies with
$n = 100, 200 ... 1000$ rules and $d = 5$ fields. We show that the
effect of the factors compounds together, by showing the original
bound $f(n, d)$, then $g(n,d)$, the bound in presence of these
factors, introducing first $allprob$, then narrow fields, and finally
$oneprob$.

As the maximum size in the presence of singletons requires us to
compute and check all partitions - an exponential number - we do not
compute the bound for such large $n$; we provide the impact of
$oneprob$ for $n = 10, 20 ... 100$.  [The other plots were similar in
shape for $n = 10, 20 ... 100$ and $n = 100, 200 ... 1000$.] We also
checked that these bounds are in fact correct by generating decision
diagrams for a large number of policies (a total of approx. two
million); the bounds were never exceeded.

\section{Related Work}

Owing to the practical importance of policies, there exists a
considerable body of work devoted to their study. A natural question
is how this paper fits into the context of this research. 

The first and most obvious area related to our work is the theory of
policies. This area includes the design of algorithms and data
structures for fast packet processing (i.e. resolution) as well as for
verification of policies. Policies have been represented as tries
\cite{Trie} and lookup-tables by Waldvogel \cite{Waldvogel} and Gupta
\cite{Gupta}. As preprocessing can be expensive, solutions have been
developed by Suri\cite{Suri} using B-trees, and by Sahni and
Kim\cite{Sahni} using red-black trees and skip lists; these solutions
allow fast update, and also perform (longest prefix) matching in
$O(\log n)$ time.

Our work not only advances the theory of policies by providing an
exact bound on the size of the policy decision diagram
\cite{LiuGMN04}, but also proposes new metrics for the complexity of a
policy, and shows why the algorithms for fast resolution and
verification are tractable in practice. Our new metrics (fieldwidth,
allprob and oneprob) make it possible to answer if a policy will have
a decision diagram of tractable size.

More generally, the analysis of policies includes the study of
anomalies \cite{HamedAM05}, inter-rule conflicts \cite{EppsteinM01},
redundancies, and so on. For example, Frantzen \cite{FrantzenKSF01}
provides a framework for understanding the vulnerabilities in a
firewall, and Blowtorch \cite{Blowtorch} is a framework to generate
packets for testing. These algorithms depend on the complexity of a
policy, as measured by the size of its decision diagram; our work
sheds light on why they are practical to use.

Finally, our work impacts the study of high-performance architecture
for fast packet resolution.  High-throughput systems, such as backbone
routers, make use of special hardware - ternary content addressable
memory\cite{TCAM}, ordinary RAM, pipelining systems \cite{Basu}, and
so on.  But these systems are very expensive, as well as limited in
the size of policies they can accommodate. Therefore, there is active
research on algorithms to optimize policies, such as Liu's TCAM
Razor\cite{TCAM}. This paper shows how, using our concept of rules in
play, we can improve upon these algorithms to create truly minimum
policies (rather than just locally minimal ones).

\section{Concluding Remarks}

In this paper, we make two contributions to the theory of policies and
their complexity.  Our first contribution is to introduce the concept
of ``rules in play'', which enables us to calculate the actual upper
bound on the size of the policy decision diagram. We also show why
this size does not grow explosively for practical policies, and
introduce new metrics for the complexity of a policy. These metrics
(oneprob, allprob and fieldwidth) are not only simple to compute (one
pass through the policy, $O(nd)$ time), but also have a dramatic
effect on the behavior of policy algorithms. Thus, our work provides
an explanation for the ``unreasonable effectiveness'' of practical
decision diagram based algorithms: their running time is not really
$O(n^d)$, but depends on our new metrics also, and practical policies
have tractable values for these new metrics.

Our second contribution is that, using rules in play, we propose the
first optimization algorithm that can produce a truly minimum-length
policy (rather than simply a minimal one, as can be found by removing
redundant rules from the policy until no redundancy remains). 

Our work on decision diagrams suggests several problems for further
study. How do our metrics influence other measures of complexity, such
as how many rules the average rule in a policy overlaps with? Can they
be further refined (for example, by taking the oneprob and allprob
field by field, rather than one measure for the whole policy)? And how
do they influence other algorithms and data structures in policy
research, such as decision-diagram compression by Bit Weaving
\cite{CRM}? For our own future work, we are focusing on the last of
these questions. We also aim to improve our new algorithm for policy
optimization, so as to make it fast enough for practical use.

\bibliographystyle{IEEEtran} \bibliography{biblio}

\end{document}